\documentclass[10pt,aps,prd,twocolumn,notitlepage,nofootinbib,floatfix,
superscriptaddress]{revtex4-1}

\usepackage{amsmath, amsthm, amssymb, amsfonts, amsbsy, mathrsfs}
\usepackage[caption=false]{subfig}

\usepackage{xcolor}
\usepackage{calligra,bm}
\usepackage{stmaryrd}
\usepackage{subfig}

\usepackage{yfonts}

\usepackage{graphicx}
\graphicspath{{./images/}}

\usepackage[hidelinks,bookmarks=true]{hyperref}
\hypersetup{pdfstartview=FitH,pdfhighlight=/O,colorlinks=false}

\begin{document}

\title{Self-collision of a portal wormhole}

\author{Justin C. Feng}
\affiliation{Centro de Astrof\'{\i}sica e Gravita\c c\~ao  - CENTRA,
Departamento de F\'{\i}sica, Instituto Superior T\'ecnico - IST,
Universidade de Lisboa - UL,
Av. Rovisco Pais 1, 1049-001 Lisboa, Portugal}
\author{Jos{\'e} P. S. Lemos}
\affiliation{Centro de Astrof\'{\i}sica e Gravita\c c\~ao  - CENTRA,
Departamento de F\'{\i}sica, Instituto Superior T\'ecnico - IST,
Universidade de Lisboa - UL,
Av. Rovisco Pais 1, 1049-001 Lisboa, Portugal}
\author{Richard A. Matzner}
\affiliation{Theory Group and Center for Gravitational Physics,
The University of Texas at Austin, Texas 78712, USA}

%\date{\today}

%-----------------------------------------------------------------------
%-----------------------------------
%-----------------
%--------
%---
%-
%
%
%-
%---
%--------
%-----------------
%-----------------------------------
%-----------------------------------------------------------------------

%=======================================================================
%-----------------------------------------------------------------------
%
%		ABSTRACT
%
%-----------------------------------------------------------------------
%=======================================================================
\begin{abstract}
We consider the self-collision of portals in classical general
relativity. Portals are wormholes supported by a single loop of negative
mass cosmic string, and being wormholes, portals have a nontrivial
topology. Portals can be constructed so that the curvature is zero
everywhere outside the cosmic string, with vanishing ADM mass. The
conical singularities of these wormholes can be smoothed, yielding a
spatial topology of $S^2 \times S^1$ with a point corresponding to
spatial infinity removed. If one attempts to collide the mouths of a
smoothed portal to induce self-annihilation, one naively might think
that a Euclidean topology is recovered, which would violate the
classical no topology change theorems. We consider a particular limit of
smoothed portals supported by an anisotropic fluid, and find that while
the portal mouths do not experience an acceleration as they are brought
close together, a curvature singularity forms in the limit that the
separation distance vanishes. We find that in general relativity, the
interaction between portal mouths is not primarily gravitational in
nature, but depends critically on matter interactions.
\end{abstract}

%-----------------------------------------------------------------------
%-----------------------------------
%-----------------
%--------
%---
%-
%
%
%-
%---
%--------
%-----------------
%-----------------------------------
%-----------------------------------------------------------------------

\maketitle

%-----------------------------------------------------------------------
%-----------------------------------
%-----------------
%--------
%---
%-
%
%
%-
%---
%--------
%-----------------
%-----------------------------------
%-----------------------------------------------------------------------

%=======================================================================
%-----------------------------------------------------------------------
%
%		INTRODUCTION
%
%-----------------------------------------------------------------------
%=======================================================================

\section{Introduction} \label{sec:intro}

A portal is a type of traversable wormhole described in
\cite{Visser1995} as a loop-based wormhole supported by a single loop of
negative mass/negative tension cosmic string. Portals can be thought of
as a limiting case of a spherical thin-shell wormhole that is flattened
to a disk. Portals may be constructed so that the geometry is flat
everywhere except at the string. The dihedral wormhole, the
two-dimensional version of a polyhedral wormhole considered in
\cite{Visser1989}, is an example of a portal (with corners). To construct
a portal one can employ the thin-shell formalism to compute the
energy-momentum tensor for the matter distribution required to hold it
open. One finds that a tremendous amount of negative mass is needed to
support portals of this type; a calculation in \cite{Visser1995}
estimates that a square-shaped dihedral wormhole \cite{Visser1989} with
a surface area of $\sim 1 \, \text{m}^2$ requires a cosmic string on the
wormhole boundary with a negative mass of the magnitude of the mass of
Jupiter to hold it open.  Portals may be obtained by taking limits of
other spacetime geometries, such as the zero-mass limit for an
appropriate analytic extension of the Kerr metric
\cite{Gibbons:2017jzk,Gibbons:2017djb,Gibbons:2016bok} (there, portals
are referred to as gates); see also \cite{Zipoy1966} which discusses
similar wormhole structures in a class of spheroidal solutions to the
vacuum Einstein equations. In this article, we adopt the terminology of
\cite{Krasnikov2003,Krasnikov2018book} for portals since they are now
widely known as such in popular culture due to examples found in media,
for instance, those found in the eponymous video games
\cite{game:Portal,*game:Portal2}.

Quantum gravitational considerations suggest that spacetime at the
Planck scale may permit fluctuations in topology
\cite{Wheeler1955,MTW,Hawking1978}, so one might expect portal-like
wormholes to form during the Planck epoch of the early universe. In the
case such wormholes are stable or metastable, cosmic inflation could
expand them to macroscopic sizes \cite{Cramer1995}. Of course, whether
portals or portal-like wormholes are stable depends on the microscopic
description of the cosmic string required to support the portal. The
locally defined mass for such a cosmic string is negative, so a
microscopic model will necessarily violate energy conditions. More
generally, it has been shown under some rather general considerations
that traversable wormholes \cite{Morris1988a} require the violation of
the null energy condition \cite{Morris1988b,Friedman1993,Hochberg1998};
see also \cite{Visser1995}. However, such a violation does not
necessarily imply that a solution is unphysical. For instance, the
avoidance of energy condition violations in the energy-momentum tensor
for matter has been treated within modified gravity theories
\cite{Rosa2018}, and it has been shown that traversable wormhole
solutions can be constructed in Einstein-Dirac-Maxwell theory, the
fermionic sector of which violates the null energy condition
\cite{Blazquez-Salcedo2021}. For a detailed discussion of energy
conditions, we refer to the review \cite{Kontou2020}.

In this work, we consider the smoothing and self-collision of portals,
assuming a simple matter model for the cosmic string. Being wormholes,
portals have a nontrivial spatial topology. Indeed, we will construct
smoothed portals and show that their topology in three spatial
dimensions is $S^2 \times S^1$ minus a point corresponding to spatial
infinity. For an axisymmetric portal configuration, one might imagine
taking a limit in which the portal mouths are brought together in a
manner in which one would naively expect to recover a Euclidean
topology. However, this would run contrary to classical no topology
change theorems \cite{Geroch1970,Lee1978}. Such a limit is
nonetheless useful for identifying and studying obstructions to topology
change in classical general relativity, and may perhaps be of interest
for studying topology-changing processes in quantum gravity.
Furthermore, an understanding of the self-interaction of portals can
(stability issues aside) shed some light on whether portals are expected
to persist after the Planck epoch of the early universe.

In Sec.~\ref{sec:portalcons}, we describe the construction of portals.
In Sec.~\ref{sec:smoothstring}, we discuss the smoothing of the conical
singularities in portals. In Sec.~\ref{sec:topology}, we carefully
illustrate the topology of smoothed portals. In
Sec.~\ref{sec:lineelement}, a line element for static smoothed
axisymmetric portals in cylindrical coordinates is constructed. In
Sec.~\ref{sec:junctioncond}, junction conditions are analyzed. In
Sec.~\ref{sec:selfinteraction}, we examine the self-collision of
portals. In Sec.~\ref{conc}, we discuss results. Throughout this
article, we consider a $3+1$ dimensional spacetime and employ the
$\left(-,+,+,+\right)$ signature, choosing units such that the
gravitational constant $G$ and the speed of light are set to one, $G=1$
and $c=1$. Unless stated otherwise, all diagrams in this article are
spatial, i.e., they describe the geometry and topology of spatial
hypersurfaces.

%=======================================================================

%-----------------------------------------------------------------------
%-----------------------------------
%-----------------
%--------
%---
%-
%
%
%-
%---
%--------
%-----------------
%-----------------------------------
%-----------------------------------------------------------------------

%=======================================================================
%-----------------------------------------------------------------------
%
%		PORTALS: CONSTRUCTION
%
%-----------------------------------------------------------------------
%=======================================================================
\section{Portals: Construction} \label{sec:portalcons}

Portals may be constructed by way of a cut-and-paste procedure along a
pair of two-dimensional disks of radius $a$ in flat three-dimensional
Euclidean space. One can imagine portals as a limiting case of a
spherical thin-shell wormhole that is flattened to a disk; the
thin-shell formalism may then be employed to compute the energy-momentum
tensor for the matter distribution required to hold it open.

\begin{figure}[!ht]
\includegraphics[width=1.0\columnwidth]{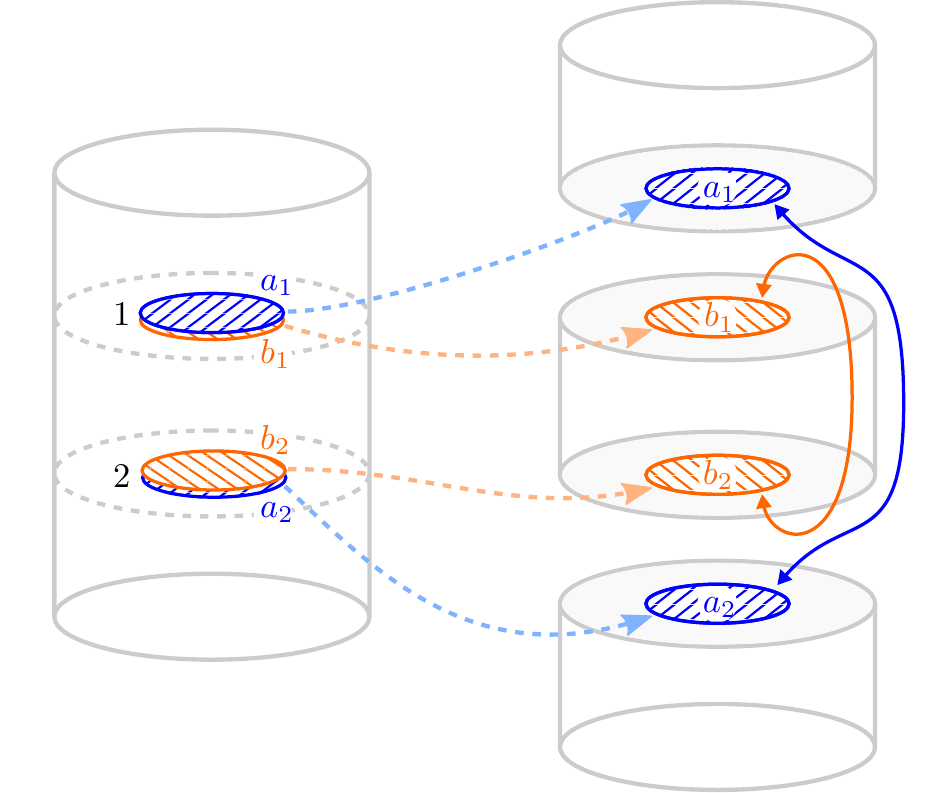}
\caption{
Gluing procedure for axisymmetric portals. The left side of the diagram
describes a cylindrical region in the three-dimensional Euclidean space
centered on the $z$-axis. Disks $1$ and $2$ are on the respective planes
defined by $z=z_0$ and $z=-z_0$, have radius $a$ and are also centered
on the $z$-axis. The right side of the diagram describes the gluing
procedure for the disks, where one performs the identifications $a_1
\leftrightarrow a_2$, and $b_1 \leftrightarrow b_2$.
}
\label{fig:Glue}
\end{figure}

\begin{figure}[!ht]
\includegraphics[width=1.0\columnwidth]{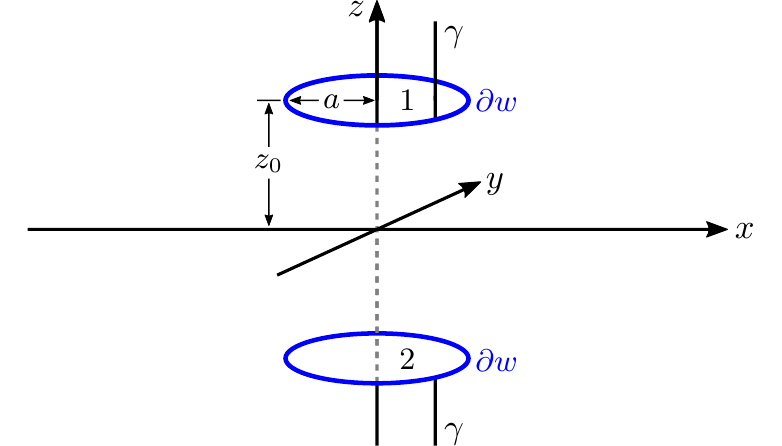}
\caption{
Diagram of a portal in three-dimensional space, assumed Euclidean. The
portal wormhole $w$ has radius $a$ and boundary denoted by $\partial w$.
The curve $\gamma$ is parallel to the $z$-axis. Observe that $\gamma$
does not pass through the region $-z_0<z<z_0$, $a\leq 1$. Though the
boundary to the wormhole mouths $\partial w$ appears in two places, it
is in fact a single surface. Here the two previous blue and orange
colors used in the identification process of the two sides of the portal
are combined into a single blue color to simplify the figure.
}
\label{fig:Portal}
\end{figure}

\begin{figure}[htp]
\includegraphics[width=1.0\columnwidth]{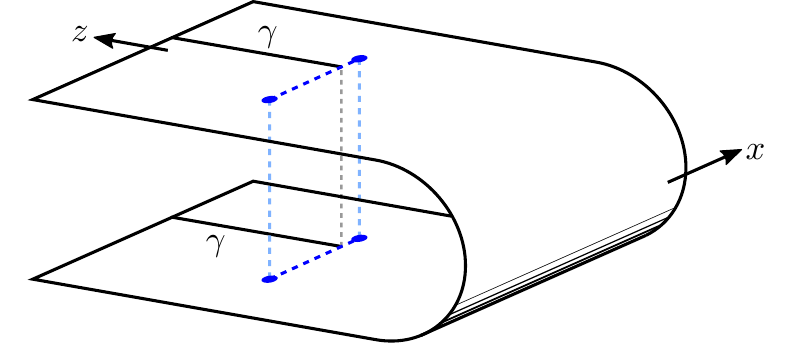}
\caption{An illustration of the $x$-$z$ plane of Fig.~\ref{fig:Portal},
embedded in a different three-dimensional Euclidean space. Here, the
vertical dotted lines have a proper distance of zero. Again, the two
previous blue and orange colors used in the identification process of
the two sides of the portal have been mixed into a single blue color.}
\label{fig:Portalxz}
\end{figure}

Let both disks be centered on the $z$-axis, with disk $1$ lying in the
plane $z=z_0$ and disk $2$ lying in the plane $z=-z_0$. The cut and
paste procedure, as illustrated in Fig.~\ref{fig:Glue}, involves gluing
the top face of disk $1$, denoted by $a_1$, to the bottom face of disk
$2$, denoted by $a_2$, and gluing the bottom face of disk $1$, denoted
by $b_1$, to the top face of disk $2$, denoted by $b_2$. The result is a
portal wormhole, i.e., a portal. Such a wormhole is illustrated in
Fig.~\ref{fig:Portal}, which also shows a curve that passes into the
bottom face $a_2$ and emerges from the top face $a_1$.  In
Fig.~\ref{fig:Portalxz} a cut of the space through $y=0$ is made and the
two-dimensional $x$-$z$ plane of the wormhole is shown embedded in a
different three-dimensional Euclidean space.  Note that to simplify the
analysis, we shall only consider cases where the identifications are
performed without twisting, meaning that a frame transported along a
straight line traveling through the wormhole does not experience any
change in orientation.

It is appropriate at this point to introduce some terminology. The
surfaces of disk $1$ and disk $2$ will be referred to as the mouths of
the wormhole $w$. The boundary of disks $1$ and $2$, which is really a
single surface of codimension two, will be denoted $\partial w$ and
referred to as the wormhole boundary.

\begin{figure}[!htb]
\subfloat[\label{subfig:4a}]{%
  \includegraphics[width=1.0\columnwidth]
  {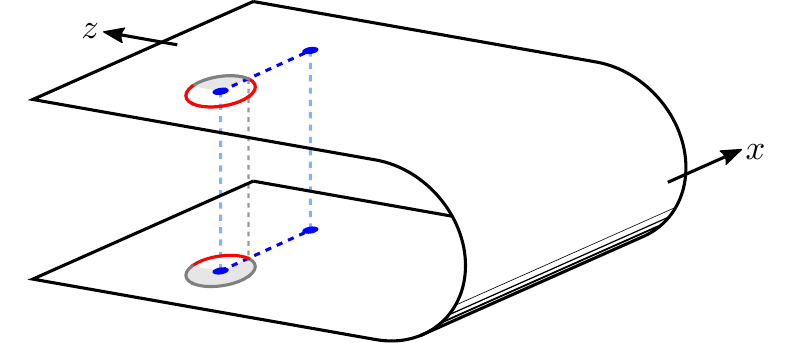}}\\
\subfloat[\label{subfig:4b}]{%
  \includegraphics[width=1.0\columnwidth]
  {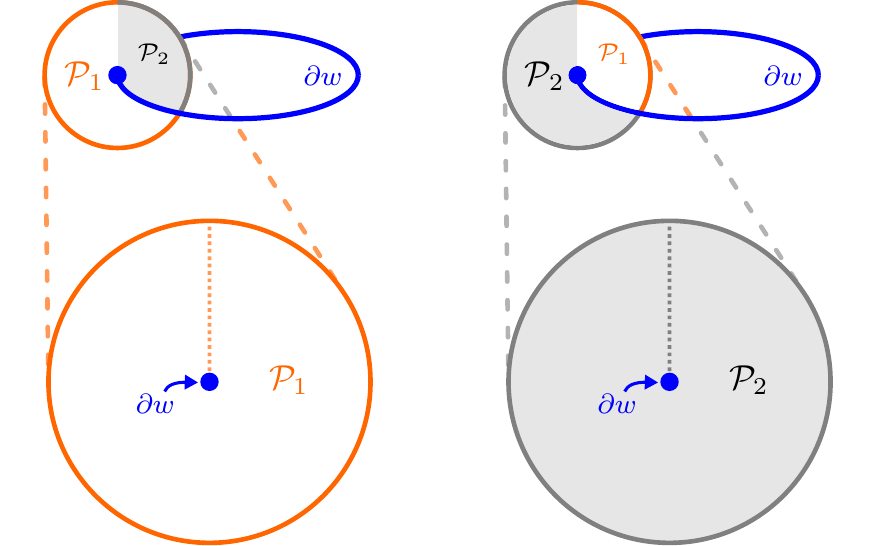}}

\caption{
Illustration of surface $\mathcal{P}$ in the neighborhood of $\partial
w$. $\mathcal{P}$ is defined to be the subset to a plane orthogonal to
the tangent vectors of $\partial w$. In (a), $\mathcal{P}$ is
illustrated in the $x$-$z$ plane as the region enclosed by the circle.
In (b), the surface $\mathcal{P}$ is a single surface divided into two
regions labeled $\mathcal{P}_\text{1}$ and $\mathcal{P}_\text{2}$. In
the bottom portion of the diagram, $\mathcal{P}_\text{1}$ and
$\mathcal{P}_\text{2}$ are illustrated as flat disks, which have a cut
along the vertical half-dotted lines---they are glued together in the
manner indicated in the top portion of (b) to form the surface
$\mathcal{P}$. Again, the two previous blue and orange colors used in
the identification process of the two sides of the portal have been
given in blue.
}
\label{fig:PortalGeom}
\end{figure}

The wormhole boundary is singular, where we define singularities in
terms of geodesic incompleteness \cite{HawkingEllis}, and in particular,
we regard a singularity to be any (limit) point or submanifold through
which one cannot extend a geodesic. The singularity here is rather mild;
it is in fact a conical singularity characterized by a surplus angle of
$2 \pi$. To see this, consider Fig.~\ref{fig:PortalGeom}, which
describes the geometry surrounding the singularity in terms of a surface
$\mathcal{P}$, which is a subset of a plane normal to the wormhole
boundary $\partial w$. In particular, $\mathcal{P}$ is defined to be a
two-dimensional surface cutting through the wormhole boundary $\partial
w$ such that the tangent vector of $\partial w$ is normal to the plane
$\mathcal{P}$. In Fig.~\ref{fig:PortalGeom}\subref{subfig:4a},
$\mathcal{P}$ is illustrated in the $x$-$z$ plane as the region enclosed
by the circle. From Fig.~\ref{fig:PortalGeom}\subref{subfig:4b} one can
infer that, excluding the singular point at $\partial w$, $\mathcal{P}$
can be decomposed into two flat disks $\mathcal{P}_1$ and
$\mathcal{P}_2$ which are each cut along a radial line and glued
together along the cut lines such that the disk $\mathcal{P}_2$ is
effectively inserted into the cut line of $\mathcal{P}_1$; the point
$\partial w$ in the surface $\mathcal{P}$ may then be characterized by a
surplus angle of $2 \pi$.


A glimpse of the self-collision problem, or the limit $z_0\rightarrow
0$, can be now advanced. At first glance, the portal appears to
annihilate itself, since it is naively expected that the wormhole
approaches Euclidean space in this limit. However, being a wormhole, the
topology of a portal is nontrivial, so the complete self-annihilation of
a portal implies a change in the topology of the manifold. Such a
process will likely be of interest for studying topology change in
quantum gravity. A natural question, the answer to which forms one of
the important topics of this article, is whether gravity encourages or
obstructs the collision between the mouths of portals supported by
smoothed cosmic strings.

%=======================================================================

%-----------------------------------------------------------------------
%-----------------------------------
%-----------------
%--------
%---
%-
%
%
%-
%---
%--------
%-----------------
%-----------------------------------
%-----------------------------------------------------------------------

%=======================================================================
%-----------------------------------------------------------------------
%
%		SMOOTHED COSMIC STRINGS
%
%-----------------------------------------------------------------------
%=======================================================================

\section{Smoothed cosmic strings} \label{sec:smoothstring}

The metric for a straight nonsingular cosmic string can be written in
cylindrical coordinates $(t,r,\varphi,\zeta)$ as $ds^2 = -dt^2 + dr^2 +
\alpha^2 \, r^2 \, d\varphi^2 + d\zeta^2$. For $\alpha$ different from
unity this metric has a conical singularity at $r=0$. To smooth this
singularity out one introduces a function $\psi=\psi(r)$ such that the
line element takes the form
\begin{equation} \label{RCM-Metric4D}
ds^2 = -dt^2 + \psi \, dr^2 + \alpha^2 \, r^2 \, d\varphi^2 + d\zeta^2 .
\end{equation}

\noindent One may read off the components of the metric tensor $g_{\mu
\nu}$ by  comparison with $ds^2 = g_{\mu \nu} dx^\mu dx^\nu$. The string
smoothing function $\psi$ must satisfy the properties,
\begin{equation} \label{RCM-SmoothingFunc}
\begin{aligned}
\lim_{r\rightarrow 0} \psi(r) &= \alpha^2\,,\quad\quad
\lim_{r\rightarrow \infty} \psi(r) &= 1 \,.
\end{aligned}
\end{equation}

\noindent For a fixed domain $0<\varphi<2\pi$, (with $\varphi=0$ and
$\varphi=2\pi$ identified), $\alpha<1$ corresponds to a line element for
a conical spacetime in the limit $\psi \rightarrow 1$ with a deficit
angle $\delta=2\pi(1-|\alpha|)$. For portals, the relevant parameter
choice is $\alpha=2$, which corresponds to a surplus angle of $2 \pi$ in
the singular limit $\psi \rightarrow 1$. The singular string is at
$r=0$; the smoothing should maintain the {\it core} of the string at
$r=0$, with possible shifts $\Delta r$ much smaller than the width of
the smoothing region.

One can evaluate the Einstein tensor $G_{\mu \nu}$ for the metric given
in Eq.~(\ref{RCM-Metric4D}) which has the following nonvanishing
components
\begin{equation} \label{RCM-EinsteinTensor}
\begin{aligned}
G_{tt}  &= -\frac{1}{2r}\frac{\partial}{\partial r}
\left[\frac{1}{\psi}\right] \\
G_{\zeta\zeta}& = \frac{1}{2r}\frac{\partial}{\partial r}
\left[\frac{1}{\psi}\right],
\end{aligned}
\end{equation}

\noindent so that $G_{\zeta\zeta}  = -G_{tt}$. Combined with the
Einstein field equations $G_{\mu \nu} = 8 \pi T_{\mu \nu}$, where
$T_{\mu \nu}$ is the energy-momentum tensor, this result suggests a
rudimentary model for a negative mass cosmic string in general
relativity: one may model a straight, negative mass cosmic string as a
negative mass anisotropic fluid with energy momentum tensor
$T_{\cdot\cdot}=\text{diag}(\rho_e,p_r,p_\varphi,p_\zeta)$ and equation
of state $p_r=0$, $p_\varphi=0$, and $p_\zeta=-\rho_e$, where
$\rho_e<0$, and with $\rho_e$ being the energy density, and $p_r$,
$p_\varphi$, and $p_\zeta$, being the stresses in the respective
directions. One might expect the equation of state $p_\zeta=-\rho_e$ to
lead to instabilities, since one might generally expect systems
containing negative energy matter to be unstable. However, these issues
of instability concern the microscopic features of the cosmic string.
For our purposes, this simple model, which we imagine to be a
coarse-grained description of a cosmic string, suffices.

%=======================================================================

%-----------------------------------------------------------------------
%-----------------------------------
%-----------------
%--------
%---
%-
%
%
%-
%---
%--------
%-----------------
%-----------------------------------
%-----------------------------------------------------------------------

%=======================================================================
%-----------------------------------------------------------------------
%
%		Topology of smoothed portals
%
%-----------------------------------------------------------------------
%=======================================================================
\section{Topology of smoothed portals} \label{sec:topology}

Now we consider the topology of a portal in a spatial slice. Since one
can smooth out the conical singularities for cosmic strings, one can
construct a manifold containing a portal that is everywhere regular. To
simplify the discussion, we describe the topology in the context of the
axisymmetric portal configuration as illustrated in
Fig.~\ref{fig:Portal}.

\begin{figure}[!ht]
\subfloat[\label{subfig:5a}]{%
\includegraphics[width=1.0\columnwidth]{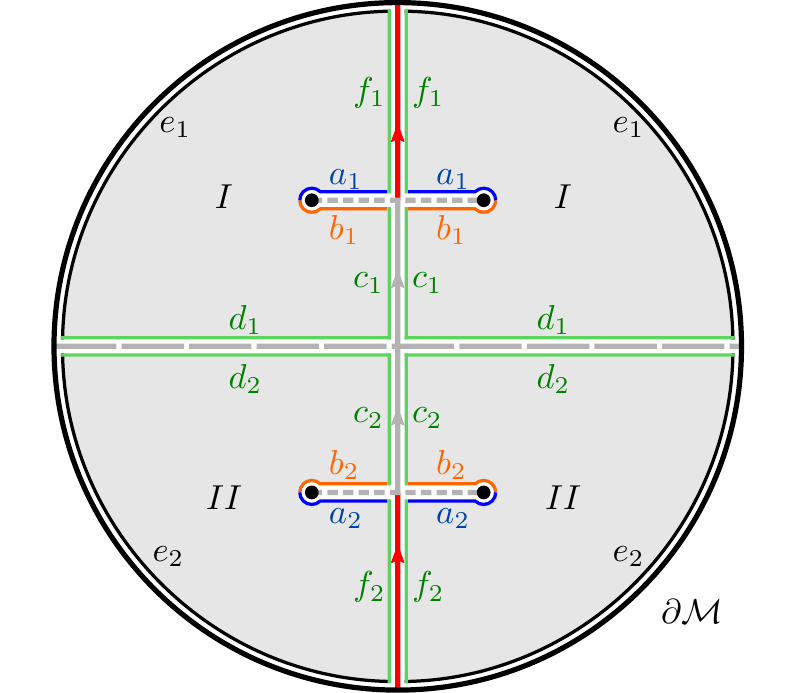}%
}\\
\subfloat[\label{subfig:5b}]{%
\includegraphics[width=1.0\columnwidth]{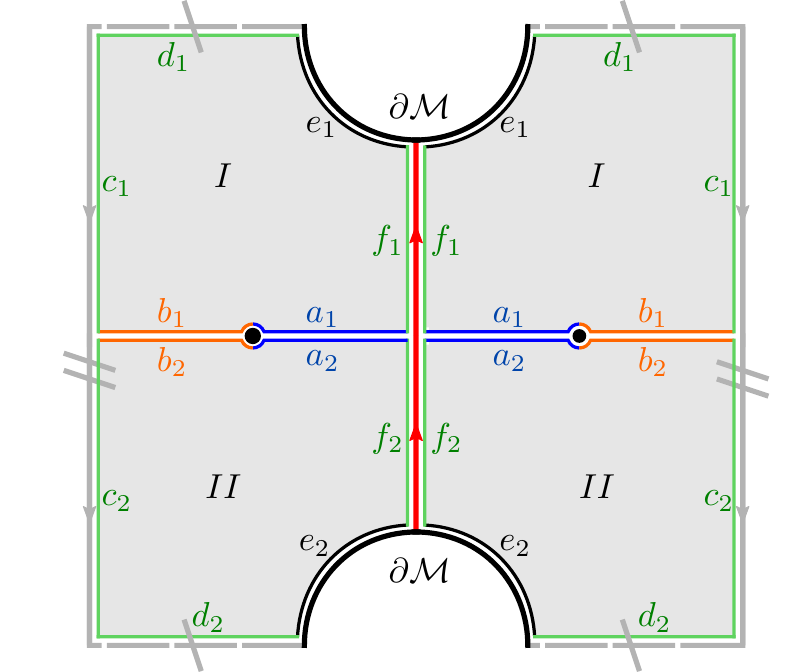}%
}
\caption{
Topology of a portal. (a) describes the compactified $x$-$z$ plane of
the axisymmetric portal of Fig.~\ref{fig:Portal}, with $\partial
\mathcal{M}$ being the boundary at $\tilde{r}=\sqrt{x^2+z^2}=\infty$.
Here, the dotted lines (representing the mouths of the wormhole) denote
cuts in the manifold, and the vertical line in the center corresponds to
the $z$-axis. (b) is an illustration of the same compactified plane,
with the vertical and horizontal sides identified. The contours
enclosing the shaded regions $I$ and $II$ form the boundaries of $I$ and
$II$, and their segments are labeled to illustrate the correspondence
between the diagrams.  The segments labeled $c_1$, $c_2$, $f_1$, $f_2$
all lie along the $z$-axis, and the segments $e_1$ and $e_2$ lie along
the boundary $\partial \mathcal{M}$. The remaining segments are glued:
$a_1$ is glued to $a_2$, $b_1$ is glued to $b_2$, and $d_1$ is glued to
$d_2$.
}
\label{fig:Topology}
\end{figure}

\begin{figure}[!ht]
\subfloat[\label{subfig:6a}]{%
\includegraphics[width=0.95\columnwidth]
{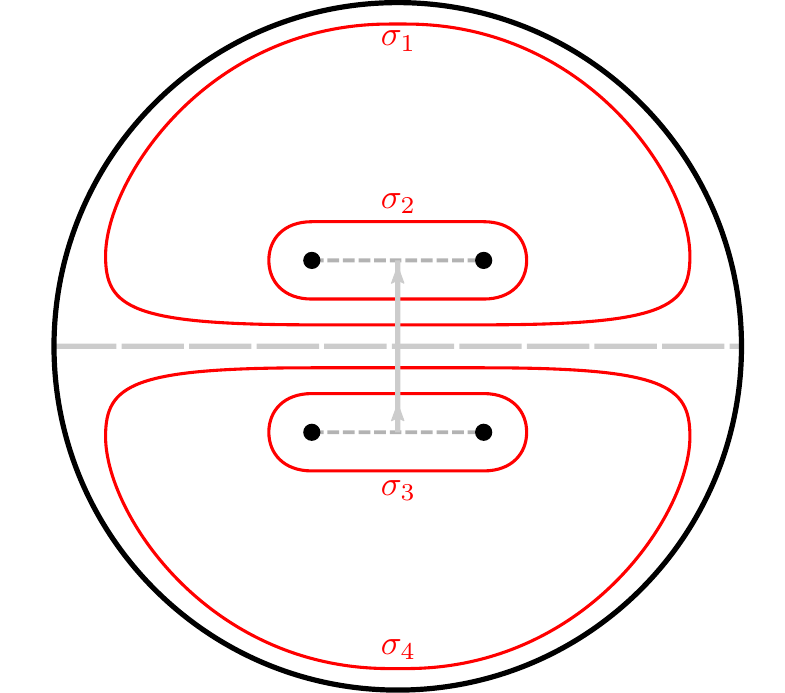}%
}\\
\subfloat[\label{subfig:6b}]{%
\includegraphics[width=0.95\columnwidth]{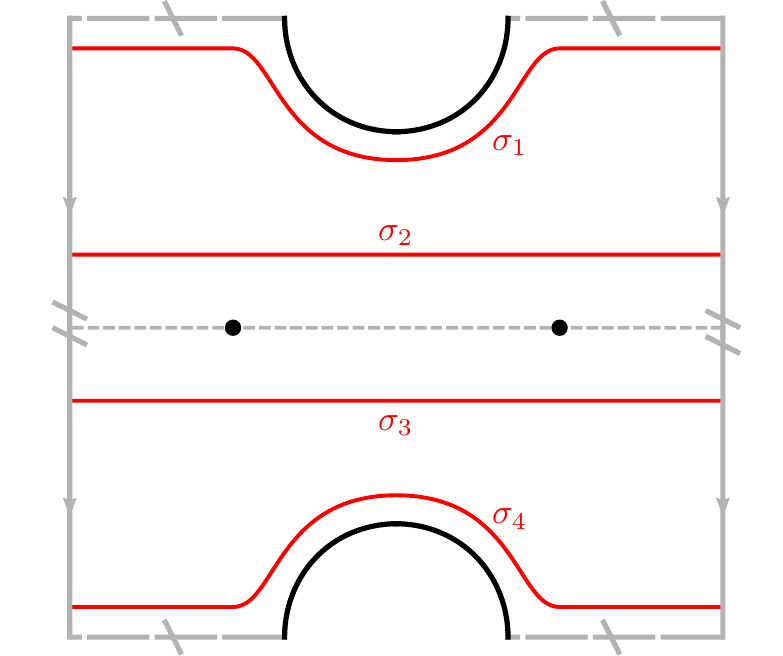}%
}
\caption{
Behavior of surfaces in Fig.~\ref{fig:Topology}. Since (a) is a
representation of an axisymmetric configuration, one can imagine
rotating (a) about a vertical line in the center so that the contours
here represented by $\sigma_1$, $\sigma_2$, $\sigma_3$, and $\sigma_4$,
all describe surfaces with topology $S^2$. Since the vertical gray lines
in (b) are identified [these correspond to the vertical gray line in the
center of (a)], one can then infer that the topology is $S^2 \times S^1$
minus a point or minus a hole.
}
\label{fig:TopologyContours}
\end{figure}

Since we consider axisymmetric portals, the spatial topology can be
represented as a slice along the $x$-$z$ plane which upon
compactification is illustrated in Fig.~\ref{fig:Topology}. In
Fig.~\ref{fig:Topology}\subref{subfig:5a} a straightforward presentation
of the compactified slice is shown; note that this diagram has cuts,
indicated by the dotted lines between the contours $a_i$ and $b_i$, $i
\in \{ 1,2 \}$. In Fig.~\ref{fig:Topology}\subref{subfig:5b} a
different, more elaborate but interesting presentation of the
compactified manifold without cuts is shown. That Figs.
\ref{fig:Topology}\subref{subfig:5a} and
\ref{fig:Topology}\subref{subfig:5b} describe the same manifold can be
seen by recognizing that regions $I$ and $II$ in
Fig.~\ref{fig:Topology}\subref{subfig:5a} are homeomorphic to their
respective counterparts in Fig.~\ref{fig:Topology}\subref{subfig:5b},
and that the labeled segments of the boundaries enclosing regions $I$
and $II$ are glued together in the same way in Figs.
\ref{fig:Topology}\subref{subfig:5a}
and~\ref{fig:Topology}\subref{subfig:5b}.

Figure \ref{fig:TopologyContours} is a schematic image of Fig. 5, and is
helpful for understanding the topology of the portal. From it, one can
infer the three-dimensional topology by recognizing that the possible
curves $\sigma^i$ which could be drawn on
Fig.~\ref{fig:Topology}\subref{subfig:5a} and are displayed
explicitly in Fig.~\ref{fig:TopologyContours}\subref{subfig:6a}, indeed
represent surfaces with the topology of a two-sphere $S^2$.  Moreover,
the same curves that could be drawn on
Fig.~\ref{fig:Topology}\subref{subfig:5b} are displayed in
Fig.~\ref{fig:TopologyContours}\subref{subfig:6b}, now appearing as
horizontal lines. The identifications indicate a topology $S^2 \times
S^1$ with a point (or sphere-shaped hole) corresponding to spatial
infinity $\partial \mathcal{M}$, which is represented by the
semicircular arcs in Figs.~\ref{fig:Topology}\subref{subfig:5b}
and~\ref{fig:TopologyContours}\subref{subfig:6b}.

As an aside, we remark on the relationship of the construction performed
here to the Deutsch-Politzer (DP) time machine, which is constructed by
way of a similar cut-and-paste procedure in $n$-dimensional spacetime
\cite{Deutsch1991,Politzer1992,Chamblinetal1994,Yurtsever1995}. In
particular, the 2+1 counterpart of the DP spacetime can be easily
visualized by replacing the $z$-axis in Fig.~\ref{fig:Portal} with the
$t$-axis. As a further aside, one might also imagine an interesting
variant of the DP construction in which the mouths are placed side by
side. In particular, one can in four-dimensional Minkowski space perform
a cut-and-paste procedure on the $t=0$ slice along two nonoverlapping
regions $s_1$ and $s_2$, each bounded by a two-sphere of the same
radius, such that timelike and null curves passing through $s_1$ from
$t<0$ emerge through $s_2$ at $t>0$. Such a construction is homeomorphic
to the DP spacetime (though having a different causal structure, lacking
closed timelike curves), and might be thought of as describing a form of
teleportation in which two regions of space effectively switch
places---for this reason, it may be appropriate to refer to them as
\textit{teleporters}. The topology of DP spacetimes (and teleporters,
since they are homeomorphic) in $n$ dimensions is $S^{n-1} \times S^1$
minus a point corresponding to infinity
\cite{Chamblinetal1994,Yurtsever1995}, which for $n=3$ is the same as
the spatial topology of the portals considered here. However, it was
also shown in \cite{Chamblinetal1994,Yurtsever1995} that one cannot
construct an everywhere smooth metric of Lorentzian signature on these
spacetimes, in contrast to the smooth Euclidean signature metrics we
will construct in this paper. Moreover, though the spacelike
quasiregular singularities of teleporters and DP spacetimes resemble
those of conical singularities, they differ in that spacelike
quasiregular singularities can significantly alter the causal structure
of spacetime in the vicinity of the singularity - see in particular
Fig.~3 of \cite{EllisSchmidt1977}. For this reason, one cannot regard
teleporters and DP spacetimes as limits of smooth spacetimes, making
them difficult to study within the framework of classical general
relativity. Note that though one might imagine that the formation of
spacelike quasiregular singularities is forbidden by some physical
principle, the formulation of such a principle (beyond excluding by fiat
those singularities) can be a rather subtle matter, as discussed in
\cite{Krasnikov2009b,Krasnikov2016}.

%=======================================================================

%-----------------------------------------------------------------------
%-----------------------------------
%-----------------
%--------
%---
%-
%
%
%-
%---
%--------
%-----------------
%-----------------------------------
%-----------------------------------------------------------------------

%=======================================================================
%-----------------------------------------------------------------------
%
%		STATIC LINE ELEMENT FOR PORTAL GEOMETRIES
%
%-----------------------------------------------------------------------
%=======================================================================
\section{Static line element for portal geometries}
\label{sec:lineelement}

Before a proper analysis of the self-interaction and self-collision of a
portal can be conducted, it is appropriate to specify coordinates which
are simple in Fig.~\ref{fig:Topology}\subref{subfig:5a} and regular
everywhere in Fig.~\ref{fig:Topology}\subref{subfig:5b} except for a
surface of codimension one, which is dealt with later when transforming
to cylindrical coordinates. To do this, we consider the configuration
illustrated in Fig.~\ref{fig:Portal} in the singular limit where the
manifold is everywhere flat except for the conical singularity at the
wormhole boundary $\partial w$, i.e., the string core. Assume $\partial
w$ is circular, with a radius $a$, centered on the $z$-axis at $\pm
z_0$. It is natural to consider as a starting point oblate spheroidal
coordinates in the $z>0$ region, with $\partial w$ as the focal ring. In
ordinary Euclidean space, spheroidal coordinates centered on $\partial
w$ at $z=+z_0$ are related to Cartesian coordinates in the following
manner:
\begin{equation} \label{RCM-SpheroidalCoords}
  \begin{aligned}
    x &= a \cosh\mu \cos\nu \cos\phi, \\
    y &= a \cosh\mu \cos\nu \sin\phi, \\
    z &= z_0 + a \sinh\mu \sin\nu.
  \end{aligned}
\end{equation}

\noindent A reflection about the $z=0$ plane yields a coordinate system
that is regular everywhere except for a measure zero set which includes
the focal rings at $\partial w$ and the $z=0$ plane, where the
coordinates fail to be smooth. One can then apply junction conditions to
deal with the nonregularity of the coordinate system at $z=0$. The line
element for Minkowski spacetime in spheroidal
coordinates takes the form
\begin{equation} \label{RCM-LineElemSpheroidalCoords}
  \begin{aligned}
    ds^2 =& - dt^2 + \frac{1}{2} a^2
    \left(\cosh 2 \mu - \cos 2 \nu \right)
    \left( d\mu^2 + d\nu^2 \right) \\
    &+a^2 \, \cosh^2\mu \cos^2\nu \, d\phi^2 ,
  \end{aligned}
\end{equation}
A second order expansion about $\mu=0$, $\nu=0$ of the second term
yields $\frac{1}{2} a^2 \, \left(\cosh 2 \mu -\cos 2 \nu \right) \approx
a^2 \left(\mu^2 + \nu^2 \right)$. Similarly expanding the  third term
yields $a^2 \, \cosh^2\mu \cos^2\nu \approx a^2 \left(1 + \mu^2 - \nu^2
\right) $ with $\approx$ denoting the expansion to second order. Note
that $\mu=0$, $\nu=0$ corresponds to the position of the focal ring. For
large $a$, one may restrict to a small angular range in $\phi$, and
neglect terms $\mu^2 d\phi^2$ and $\nu^2 d\phi^2$. This corresponds to a
limit in which one can neglect the curvature of the wormhole boundary.
In this limit, the line element has the form:
\begin{equation} \label{RCM-LineElemSpheroidalCoordsExpansionPlane}
  \begin{aligned}
    ds^2 \approx  - dt^2 + a^2 \left(\mu^2 +
    \nu^2\right)(d\mu^2+d\nu^2) + a^2 d\phi^2.
  \end{aligned}
\end{equation}
It is straightforward to verify that this is in fact a flat metric
everywhere except at the origin point $\mu=0$, $\nu=0$. It turns out
that one can recover the conical metric given in
Eq.~\eqref{RCM-Metric4D} in the limit $\psi \rightarrow 1$ for
$\alpha=2$ with the coordinate choice given by $\mu =
\sqrt{\frac{2r}{a}} \sin \varphi$ and $\nu = \sqrt{\frac{2r}{a}} \cos
\varphi$ which yields $ ds^2 \approx  - dt^2+ dr^2+4 r^2 d\varphi^2 +
a^2 d\phi^2$. This result indicates that the geometry immediately
surrounding the conical singularity of the singular portal is given by
the metric given in
Eq.~\eqref{RCM-LineElemSpheroidalCoordsExpansionPlane} if one extends
the domain of the coordinates $\mu$ and $\nu$ to negative values. It is
straightforward to work out the following differential expressions
\begin{equation} \label{RCM-DifferentialExpressions}
  \begin{aligned}
    dr &= a \left( \mu \, d\mu + \nu \, d\nu \right), \\
    {2r}\,d\varphi &= a
    \left( \nu \, d\mu - \mu \, d\nu \right),
  \end{aligned}
\end{equation}
where $r=\frac{a}{2}(\mu^2+\nu^2)$ and $\tan \varphi=\frac\mu\nu$. One
can then use the differentials in Eq.~\eqref{RCM-Metric4D} to smooth out
the conical singularities, resulting in the line element
\begin{equation} \label{RCM-LineElemExpansionSmoothed}
  \begin{aligned}
    ds^2 \approx & - dt^2 + a^2 \left(\mu^2 + \nu^2\right)
    (d\mu^2+d\nu^2) + a^2 \, d\phi^2 \\
    & + a^2 \left(\psi-1\right)(\mu \, d\mu + \nu \, d\nu)^2.
  \end{aligned}
\end{equation}
When the string smoothing function is trivial, i.e., $\psi=1$, one
recovers Eq.~\eqref{RCM-LineElemSpheroidalCoordsExpansionPlane}. We note
that here, the point $\mu=\nu=0$ corresponds to the core of the smoothed
string ($r=0$), about which the string smoothing function is
centered.

The analysis so far requires small values for $\mu$ and $\nu$ and a
restriction to a small angular range in $\phi$. However,
Eq.~\eqref{RCM-LineElemExpansionSmoothed} can be used to motivate a line
element suitable for a larger range of coordinate values. Since $\mu$ is
a hyperbolic coordinate, and $\nu$ is an angular coordinate, it is
natural to replace instances of $\mu$ and $\nu$ in the Taylor-expanded
line element \eqref{RCM-LineElemExpansionSmoothed} with the appropriate
hyperbolic and trigonometric functions. Choosing
\begin{equation} \label{RCM-rexpressionX}
  r=\frac{a}{4} \left(\cosh 2 \mu -\cos 2 \nu \right)
\end{equation}
(note that $r=0$ at the core of the string), one may then construct the
line element
\begin{equation} \label{RCM-LineElemSpheroidalCoordsSmoothed}
  \begin{aligned}
    ds^2 = & - dt^2 + \frac{1}{2} a^2
    \left(\cosh 2 \mu -\cos 2 \nu \right)
    \left( d\mu^2 + d\nu^2 \right)  \\
    & + a^2 \, \cosh^2\mu \cos^2\nu  \, d\phi^2  \\
    & + a^2 \left(\psi-1\right)(\sinh\mu \, d\mu + \sin\nu \, d\nu)^2 ,
  \end{aligned}
\end{equation}

\noindent and it can be verified that this line element reduces to
Eq.~\eqref{RCM-LineElemExpansionSmoothed} in the appropriate limits.

It is convenient to transform
Eq.~\eqref{RCM-LineElemSpheroidalCoordsSmoothed} back to Cartesian
coordinates, since the properties of the boundary surface $z=0$, which
corresponds to the constraint (recall $z_0$ is half the $z$ separation
between the portal mouths)
\begin{equation} \label{RCM-ConstraintSurface}
\sinh\mu \sin\nu = - \frac{z_0}{a} ,
\end{equation}

\noindent are of particular interest for the self-collision problem.
In fact, since the problem is axially symmetric, it is more convenient
to transform to the cylindrical coordinates $\rho$, $\phi$, $z$, where
one has the coordinate definition $\rho := a \, \cosh\mu \, \cos\nu$,
and $z$ is defined in Eq.~\eqref{RCM-SpheroidalCoords}. One can verify
that in this cylindrical coordinate system, the line element
\eqref{RCM-LineElemSpheroidalCoordsSmoothed} takes the form
\begin{equation} \label{RCM-LineElemCylCoordsSmoothed}
  \begin{aligned}
    ds^2 =&
    - dt^2 + d\rho^2 + \rho^2 \, d\phi^2 + dz^2 \\
    & + \frac{\Psi(\rho,z) - 1}{\Delta z^2 + \Delta \rho^2}
    \left[ \Delta \rho \, d\rho + \Delta z \, dz \right]^2,
  \end{aligned}
\end{equation}

\noindent where $\Psi(\rho,z) = \psi(r)$ is the string smoothing
function, with $r$ being given by
\begin{equation} \label{RCM-rexpressionXcyl}
    r=\frac{\sqrt{\left(\Delta \rho^2 + \Delta z^2\right)
    \left[(2 a+\Delta \rho )^2 + \Delta z^2\right]}}{2 a},
\end{equation}

\noindent and the following have been defined
\begin{equation} \label{RCM-DeltaCoords}
  \begin{aligned}
    \Delta \rho &:= \rho - a ,\\
    \Delta z &:= z-z_0 .
  \end{aligned}
\end{equation}

\noindent In these coordinates, the core ($r=0$) of the smoothed string
is located at $\Delta \rho=\Delta z=0$. We compute the Einstein tensor
for Eq.~\eqref{RCM-LineElemCylCoordsSmoothed} to leading order in
$a^{-1}$, assuming $\Delta \rho \ll a$, and that $\Psi(\rho,z)$ and its
derivatives are of order unity for large $a$. The nonvanishing
components of the Einstein tensor take the form:
\begin{equation} \label{RCM-EinsteinTensorLimit}
  \begin{aligned}
    G{^t}{_t} &= \mathcal{G} + \mathcal{O}(a^{-1}) \\
    G{^\rho}{_\rho} &= \mathcal{O}(a^{-1}) \\
    G{^\phi}{_\phi} &= \mathcal{G} + \mathcal{O}(a^{-1}) \\
    G{^z}{_z} &= \mathcal{O}(a^{-1}) \\
    G{^\rho}{_z} &= G{^z}{_\rho} = \mathcal{O}(a^{-1}),
  \end{aligned}
\end{equation}

\noindent where $\mathcal{G} = \mathcal{G}(\Delta \rho, \Delta z)$ is
given by:
\begin{equation} \label{RCM-EinsteinTensorFunc}
  \begin{aligned}
    \mathcal{G}
      =~&
      \frac{1}{4 \Psi ^2 \left(\Delta \rho ^2 + \Delta z^2\right)} \\
      & \times \biggl\{
      - 2 \Delta z  \left[1 + \Psi - \Delta \rho ~ \partial_\rho
      \Psi \right] \partial_z\Psi - \Delta \rho^2 \partial_z\Psi^2 \\
      & \qquad
      - \left[
        2 \Delta \rho (1 + \Psi) \partial_\rho\Psi  + \Delta z^2
	\left(\partial_\rho\Psi\right)^2 \right] \\
      & \qquad
       + 2 \Psi  \left[
       \Delta z^2 \partial^2_\rho\Psi + \Psi \Delta \rho ^2
       \partial^2_z \Psi
       - 2 \Delta \rho \Delta z \partial_z \partial_\rho\Psi
       \right]
       \biggr\}.
  \end{aligned}
\end{equation}
\noindent As expected, we recover the Einstein tensor of
Eq.~\eqref{RCM-EinsteinTensor} in the limit of large $a$, so in this
limit, the model of the cosmic string as an anisotropic fluid with
equation of state $p_\zeta=-\rho_e$ applies. Of course, for finite $a$,
when the curvature of the cosmic string becomes significant, a more
complicated matter model will be needed.

%=======================================================================

%-----------------------------------------------------------------------
%-----------------------------------
%-----------------
%--------
%---
%-
%
%
%-
%---
%--------
%-----------------
%-----------------------------------
%-----------------------------------------------------------------------

%=======================================================================
%-----------------------------------------------------------------------
%
%		JUNCTION CONDITONS
%
%-----------------------------------------------------------------------
%=======================================================================

\section{Junction conditions} \label{sec:junctioncond}

%=======================================================================
%-----------------------------------------------------------------------
%
%		THIN SHELL AT Z=0
%
%-----------------------------------------------------------------------
%=======================================================================
\subsection{Thin shell at $z=0$}

The line element given in Eq.~\eqref{RCM-LineElemCylCoordsSmoothed} is
only valid for $z>0$. However, we can construct a metric for the $z < 0$
region simply by reflecting in the plane $z=0$, which is depicted by the
shaded region in Fig. \ref{fig:PortalCollisions}. In general, such a
procedure will create a thin shell at the $z=0$ surface. For singular
(non-smoothed) strings, the space is 3-flat, except at the locations of
the strings. Thus although there might be a coordinate discontinuity at
the $z=0$ plane, there would be no geometrical singularity. However, the
situation is different for smoothed portals, since the smoothing of the
string produces a ``fattened" matter distribution which can extend to
the $z = 0$ surface.

To see that a thin shell is created under a reflection about $z=0$, we
employ the thin shell formalism \cite{Israel1966,Poisson} to compute the
surface stress tensor at $z=0$ (here the indices $i,j$ correspond to the
coordinates $t,\rho,\phi$):
\begin{equation} \label{SurfaceEMT}
\tau_{ij} = -\frac{1}{8 \pi} \left( [K_{ij}] - [K] \gamma_{ij} \right),
\end{equation}

\noindent where $K_{ij}$ is the extrinsic curvature tensor of the
surface $z=0$, which is given by the expression $K_{ij} = \frac{1}{2
\underline{\alpha}}\left(\frac{\partial \gamma_{ij}}{\partial z} - D_i
\beta_j-D_j \beta_i \right)$, where $D_i$ is the surface covariant
derivative compatible with the induced metric $\gamma_{ij}=g_{ij}$,
$\beta_i:=g_{zi}$, $\underline{\alpha}=\sqrt{|\frac{g}{\gamma}|}$, with
$g$ and $\gamma$ being the determinants of the respective metric tensors
$g_{\mu \nu}$ and $\gamma_{ij}$, and the brackets $[A]:=A_{+}-A_{-}$
denote a jump in the extrinsic curvatures. The extrinsic curvatures are
defined such that the unit normal vectors point in the same direction
across the surface. Due to the symmetry in the problem, a nonvanishing
extrinsic curvature will generally lead to a nonvanishing jump so that
$[K_{ij}]=2K_{ij}$ and $[K] = 2 \, K$, resulting in a nonvanishing
surface energy-momentum tensor $\tau_{ij}$. A nonvanishing $\tau_{ij}$
describes the surface stress-energy of a thin shell at the $z=0$
surface, as illustrated in Fig. \ref{fig:PortalCollisions}.

Given the line element Eq.~\eqref{RCM-LineElemCylCoordsSmoothed} for
$z>0$, and demanding that the geometry be symmetric about $z=0$, the
nonvanishing components of the surface energy-momentum tensor [given by
\eqref{SurfaceEMT}] take the form
\begin{equation} \label{SurfaceEMTComponents}
\begin{aligned}
\tau{^t}{_t}
      = & \tau{^\rho}{_\rho} + \tau{^\phi}{_\phi}\,,
      \\
\tau{^\rho}{_\rho}
      = &
      \frac{\Delta \rho  (\Psi -1) z_0}
      {4 \pi (a + \Delta \rho )
      \sqrt{
            \Psi \left(\Delta \rho ^2+z_0^2\right)
            \left(\Delta \rho ^2 \Psi +z_0^2\right)}}\,,
      \\
\tau{^\phi}{_\phi}
      = &
      \frac{1}{
        8 \pi \sqrt{\Psi \left(\Delta \rho ^2+z_0^2\right)}
        \left(\Delta \rho ^2 \Psi +z_0^2\right)^{3/2}}
        \biggl\{ 2 (\Psi -1) z_0^3\\
      &
        +\Delta \rho z_0 \left(\Delta \rho ^2 (\Psi +1)+2 z_0^2\right)
        \partial_\rho \Psi \\
      &
        +\Delta \rho ^2 \left(\Delta \rho ^2 \Psi +z_0^2\right)
        \partial_z \Psi
        \biggr\}
      ,
\end{aligned}
\end{equation}

\noindent where $\Psi$ and its derivatives are evaluated at $z=0$, and
mixed index components are presented due to their simplicity. The final
expression for the components of $\tau{^i}{_j}$ depends on the choice of
a string smoothing function, and vanishes where $\Psi=1$ and $\partial_z
\Psi= \partial_\rho \Psi = 0$, outside the smoothing region. We note
that $\tau{^\rho}{_\rho}$  vanishes regardless, in the large $a$ limit.

%=======================================================================
%-----------------------------------------------------------------------
%
%		JUNCTION CONDITIONS AT Z=0
%
%-----------------------------------------------------------------------
%=======================================================================
\subsection{Junction conditions at $z=0$}

The shell at $z = 0$ is somewhat artificial, as it is the consequence
of a reflection about the $z = 0$ plane. To eliminate this shell, we
impose junction conditions, which amounts to the demand that
$\tau{^i}{_j}=0$.  These junction conditions will lead to boundary
conditions on the string smoothing function $\Psi$ and its derivatives
at $z=0$. Earlier, we saw that $\tau{^i}{_j}=0$ if $\Psi=1$ and
$\partial_z \Psi= \partial_\rho \Psi = 0$, but here we show that for
finite $a$, these boundary conditions follow from
$\tau{^i}{_j}=0$. From Eq.~\eqref{SurfaceEMTComponents}, we note that
if $\tau{^\rho}{_\rho}=0$ and $\tau{^\phi}{_\phi}=0$, then
$\tau{^t}{_t}=0$. The vanishing of $\tau{^\rho}{_\rho}$ (assuming
finite $a$) requires $\Psi|_{z=0} = 1$, which in turn implies
$\partial_\rho \Psi|_{z=0} = 0$. Under these conditions, the vanishing
of $\tau{^\phi}{_\phi}$ requires $\partial_z \Psi|_{z=0} = 0$.  Thus,
we can eliminate the shell by requiring the boundary conditions $\Psi
= 1$, $\partial_\rho \Psi =0$, and $\partial_z \Psi = 0$ at the $z=0$
surface.

In the large $a$ limit, one can see from
Eqs.~\eqref{SurfaceEMTComponents} that $\tau{^\rho}{_\rho} \rightarrow
0$, so perhaps one can in this limit relax the condition $\Psi=1$ at
$z=0$. From Eq.~\eqref{SurfaceEMTComponents}, we find that the condition
$\tau{^t}{_t}=\tau{^\phi}{_\phi}=0$ yields the following differential
equation for $\Psi|_{z=0}$:
\begin{equation} \label{Psirderiv}
\partial_\rho \Psi =
- \frac{
        \left(\Delta \rho^4 \Psi + \Delta \rho^2 z_0^2\right)
        \partial_z \Psi
        +
        2 (\Psi -1) z_0^3
       }
       {
        2 \Delta \rho  z_0^3 + \Delta \rho^3 (\Psi +1) z_0
       } ,
\end{equation}

\noindent with the understanding that all quantities here are evaluated
at $z=0$ and are at most functions of $\rho$. The right hand side of
Eq.~\eqref{Psirderiv} can be expanded in $\Delta \rho$, under the
assumption that $\Psi$ and $\partial_z \Psi$  do not diverge at $\Delta
\rho = 0$, yielding:
\begin{equation} \label{PsirderivX}
\partial_\rho \Psi = \frac{1-\Psi}{\Delta \rho} + O(\Delta \rho).
\end{equation}

\noindent If we require $\partial_\rho \Psi$ to be zero or finite at
$z=0$ and $\Delta \rho = 0$, then $(\Psi - 1)|_{z=0} \propto \Delta
\rho$, and it follows that at $z=0$ and $\Delta \rho = 0$, one must have
$\Psi \rightarrow 1$.

%=======================================================================
%-----------------------------------------------------------------------
%
%		EXAMPLE SMOOTHING FUNCTION
%
%-----------------------------------------------------------------------
%=======================================================================
\subsection{Example smoothing function}
It is not too difficult to construct a string smoothing function
satisfying the finite $a$ boundary conditions $\Psi = 1$, $\partial_\rho
\Psi = \partial_z \Psi = 0$ at the $z=0$ surface. As an example, one may
start with the following smoothing function $\psi$ \cite{Flachi2019}:
\begin{equation} \label{SmoothingFunction}
      \psi
      =
      \frac{
      r^2 + \overline{\alpha}^{\,2} \, \varepsilon^2
      }{
      r^2 + \varepsilon^2
      } ,
\end{equation}

\noindent where $\overline{\alpha}$ has the value $2$ for a smoothed
cosmic string with surplus angle $2 \pi$ at large $r$, and $\varepsilon$
is roughly the thickness of the smoothed cosmic string. From $\psi$, one
can construct a string smoothing function $\Psi$ satisfying the boundary
conditions by promoting the quantity $\overline{\alpha}$ to a function
of $z$ of the form:
\begin{equation} \label{ualphafunc}
\begin{aligned}
      \overline{\alpha}\left(\frac{z}{z_0}\right) =
      \alpha
      +
      \left(1-\alpha\right)
      \frac{\Omega(\frac{z}{z_0})}{\Omega(0)},
\end{aligned}
\end{equation}

\noindent where $\alpha=2$ for the portal geometry, and the function
$\Omega\left(x\right)$ satisfies the properties
\begin{equation} \label{OmegaProperties}
\begin{aligned}
& \Omega'(0) = \Omega(1) = 0, \\
& \Omega(0) \neq 0,\\
& \Omega\left(\infty\right) = 0.
\end{aligned}
\end{equation}

\noindent These properties are constructed to ensure that
$\bar{\alpha}=2$ at $\rho=a$ and $z=z_0$; this condition is needed to
avoid an additional conical singularity.

%=======================================================================

%-----------------------------------------------------------------------
%-----------------------------------
%-----------------
%--------
%---
%-
%
%
%-
%---
%--------
%-----------------
%-----------------------------------
%-----------------------------------------------------------------------

%=======================================================================
%-----------------------------------------------------------------------
%
%		DYNAMICAL PORTALS AND SELF COLLISION
%
%-----------------------------------------------------------------------
%=======================================================================
\section{Dynamical portals and self-collision}
\label{sec:selfinteraction}

\begin{figure}
\includegraphics[width=1.0\columnwidth]{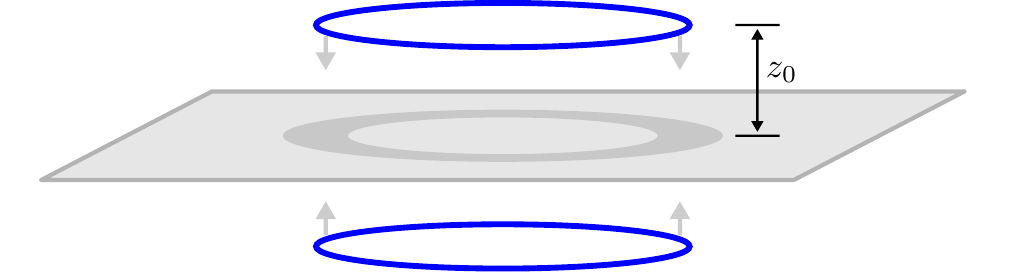}
\caption{An illustration of the self-collision problem. The $z=0$ plane
has been shaded in gray. Equation \eqref{SurfaceEMTComponents} indicates
that a generic smoothing of the string (defined by a choice for $\Psi$)
will generate a thin shell at the $z=0$ plane. For a smoothed string of
a given thickness, one expects the energy of the shell to be
concentrated in a circular strip (indicated in dark gray) centered at
$\rho=a$.}
\label{fig:PortalCollisions}
\end{figure}

%=======================================================================
%-----------------------------------------------------------------------
%
%		THE SELF-COLLISION PROBLEM
%
%-----------------------------------------------------------------------
%=======================================================================
\subsection{The self-collision problem}

The self-collision problem for an axisymmetric portal can be understood
with the help of Fig.~\ref{fig:PortalCollisions}. Here, we consider
symmetry about $z=0$. As remarked earlier, the self-collision problem
refers to a process in which the portal mouths are brought together, and
corresponds to the limit $z_0 \rightarrow 0$. It may be useful to
provide a conceptualization of the self-collision problem in the context
of Fig.~\ref{fig:Topology}. One might imagine the self-collision of a
portal as a process that brings together a portion of the portion of the
top and bottom faces of Fig.~\ref{fig:Topology}\subref{subfig:5b}. In
particular, the outermost portions of the segments $d_1$ and $d_2$ in
Fig.~\ref{fig:Topology}\subref{subfig:5b} are pinched together with
segments $b_1$ and $b_2$ to make the left and right edges (segments
$c_1$ and $c_2$) disappear. Alternatively, the distances, as defined by
the spatial metric, along a subclass of curves that pass through the
contours $b_1$, $b_2$ vanish.

To study the self-collision problem, we will consider a dynamical portal
geometry in the large $a$ limit and solve the Einstein field equations
in $3+1$ form in the vicinity of an initial slice satisfying a
generalization of the line element in
Eq.~\eqref{RCM-LineElemCylCoordsSmoothed}. The goal of this calculation
is to determine the acceleration of the portal mouths for a given set of
initial data. For the matter model, we consider an anisotropic fluid
with no stresses in the direction perpendicular to the length along the
string, as indicated in Eq.~\eqref{RCM-EinsteinTensor}; in doing so, we
ignore matter interactions in the directions perpendicular to the string
and highlight gravitational interactions between the portal mouths.

%=======================================================================
%-----------------------------------------------------------------------
%
%		TIME DEPENDENT METRIC
%
%-----------------------------------------------------------------------
%=======================================================================
\subsection{Time dependent metric}
Here, we consider the large $a$ limit. To simplify the notation, we make
the replacements $\rho \rightarrow a$ and $\Delta \rho \rightarrow
\bar{\rho}$. Without loss of generality, one may restrict to Gaussian
normal coordinates in a neighborhood of the $t=0$ slice. We generalize
the line element in Eq.~\eqref{RCM-LineElemCylCoordsSmoothed} to the
form:
\begin{equation} \label{RCM-LineElemCylCoordsSmoothedTimeDep}
  \begin{aligned}
    ds^2 = & -dt^2 + d\bar{\rho}^2 + a^2 \, d\phi^2 + dz^2
    + \frac{\Psi - 1}{\chi^2 \, \Delta z^2 + \bar{\rho}^2} \\
    & \times \biggl[ \lambda^2 \, d\rho^2
    + 2 \chi \, \bar{\rho} \, \Delta z \, d\bar{\rho} \, dz
    + \chi^2 \, \Delta z^2 \, dz^2 \biggr],
  \end{aligned}
\end{equation}

\noindent where now $\Psi=\Psi(\bar{\rho},z,t)$,
$\chi=\chi(\bar{\rho},z,t)$, and $\lambda=\lambda(\bar{\rho},z,t)$ are
smooth functions of $\bar{\rho}$ and $z$. To ensure that the geometry is
locally flat at the core of the string $\bar{\rho}=0$, $z=z_0$, we
require that $\Psi(0,z_0,0)=4$, $\chi(0,z_0,0)=1$, and
$\lambda(\bar{\rho},z_0,0)=\bar{\rho}$; notice that the line element
reduces to the form of Eq.~\eqref{RCM-LineElemCylCoordsSmoothed} at the
core of the string $\bar{\rho}=\Delta z=0$. It is also appropriate to
choose initial data such that $\dot{\Psi}=\dot{\chi}=\dot{\lambda}=0$ at
the core of the string, so the locally flat condition is maintained, at
least to first order in time. To simplify the analysis, we will choose
initial data such that at $\bar{\rho} = 0$, the quantity $\Psi$ has a
vanishing first derivative with respect to $\bar{\rho}$.

%=======================================================================
%-----------------------------------------------------------------------
%
%		GEODESIC DISTANCE BETWEEN PORTAL MOUTHS
%
%-----------------------------------------------------------------------
%=======================================================================
\subsection{Geodesic distance between portal mouths} \label{sec:geodist}

Here, we consider the length of a spatial geodesic connecting the cores
of the smoothed strings supporting the portal mouths. The core of the
string is defined to be $\bar \rho =0,\Delta z=0$ for $z>0$, and the
symmetric statement for $z<0$. For the spatial part ($dt=0$) of the line
element in Eq.~\eqref{RCM-LineElemCylCoordsSmoothedTimeDep}, one may
verify that for a tangent vector $v=(0,0,v^z)$, $^{(3)}\Gamma^i_{jk}v^j
v^k \propto v^i$ at $\bar{\rho}=0$ provided that $\partial_{\bar{\rho}}
\Psi|_{{\bar{\rho}}=0} = 0$, which follows our choice of initial data
along $\bar{\rho}=0$. Thus, the line $\bar{\rho}=0$,
$\phi=\text{constant}$ is a spatial geodesic. Along this geodesic, the
spatial part of the line element
\eqref{RCM-LineElemCylCoordsSmoothedTimeDep} simplifies to $ds^2 =\Psi
\, dz^2$, so that the length $L$ of the geodesic connecting the centers
of the smoothed strings is given by the integral:
\begin{equation} \label{ProperDistance}
L = 2 \int_0^{z_0} \sqrt{\Psi} ~ dz = L_0 + V_0 ~ t
    + \frac{A_0}{2} ~ t^2  + O(t^3).
\end{equation}

\noindent The initial distance $L_0$, the initial velocity $V_0$ and the
initial acceleration $A_0$ may be obtained by expanding $\Psi$ in $t$,
with the following result:
\begin{equation} \label{LenVelAcc}
\begin{aligned}
L_0 &:= 2 \int_0^{z_0} \sqrt{\Psi_0} ~ dz \,,\\
V_0 &:= \int_0^{z_0} \frac{\dot{\Psi}_0}{ \sqrt{\Psi_0}} ~ dz\,, \\
A_0 &:= \int_0^{z_0} \frac{2 \Psi_0  \ddot{\Psi }_0-\dot{\Psi }_0^2}
        {2\Psi _0^{3/2}} ~ dz \,,
\end{aligned}
\end{equation}

\noindent where the subscripts $0$ denote evaluation at $t=0$, overdots
denote derivatives with respect to $t$, and all quantities are
understood to be evaluated at $\bar{\rho}=0$. Since we are working in
Gaussian normal coordinates, the velocity and acceleration are measured
according to observers aligned with $\frac{\partial\;}{\partial t}$.
This result indicates that if $\partial_{\bar{\rho}}
\Psi|_{{\bar{\rho}}=0} = 0$ is assumed, one only needs to know $\Psi$
and its first and second time derivatives at $\bar{\rho}=0$ to obtain
the acceleration $A_0$ ($L_0$ and $V_0$ are determined by the initial
data).

%=======================================================================
%-----------------------------------------------------------------------
%
%		ANISOTROPIC FLUID
%
%-----------------------------------------------------------------------
%=======================================================================
\subsection{Anisotropic fluid}
For the matter model, we consider an anisotropic fluid with an
energy-momentum tensor of the form:
\begin{equation} \label{RCM-AnisotropicFluidEMT}
  \begin{aligned}
    T_{\mu \nu} = \rho_u \left( u_\mu u_\nu - w_\mu w_\nu \right)
  \end{aligned}
\end{equation}

\noindent where $\rho_u$ is the rest frame energy, $u^\mu$ is a unit
timelike vector and $w^\mu$ is a unit spacelike vector. We choose them
to have the following form:
\begin{equation} \label{RCM-AnisotropicFluidVectors}
  \begin{aligned}
    u &= \left( u^t, u^{\bar{\rho}}, 0, u^z \right) \\
    w &= \left( 0, 0, \frac1a, 0 \right)
  \end{aligned}
\end{equation}

\noindent where the component $u^t$ of the four-velocity is fixed by the
normalization condition. The initial conditions for $u^{\bar{\rho}}$ and
$u^z$ will be discussed in the next section.

%=======================================================================
%-----------------------------------------------------------------------
%
%		3+1 EQUATIONS AND THEIR SOLUTION
%
%-----------------------------------------------------------------------
%=======================================================================
\subsection{3+1 equations and their solution} \label{sec:admsol}

We will work in the $3+1$ formalism, assuming Gaussian normal
coordinates which correspond to the conditions $g_{tt}=-1$ and
$g_{ti}=0$. Here, the spatial metric will be denoted
$\gamma_{ij}=g_{ij}$ to avoid confusion. In Gaussian normal coordinates,
the $3+1$ decomposition
\cite{Alcubierre,BaumgarteShapiro,Gourgolhoun3+1} of the Einstein field
equations takes the form:
\begin{equation} \label{YADMTimeEvolutionEquationsGNC1}
\dot{\gamma}_{ij} = 2 \, K_{ij}\,,
\end{equation}
\vskip -0.99cm
\begin{equation} \label{YADMTimeEvolutionEquationsGNC2}
\dot{K}_{ij} = 2 K_{i k}K{_j}{^k}-K K_{ij} - {^3}{R}_{ij}
                - \kappa \left[\frac{1}{2} (S - \rho_m) \gamma_{ij}
                - S_{ij} \right],
\end{equation}
\begin{equation} \label{YADMMomentumConstraint}
D_{k }\left(K{_i}{^k}-\gamma{_i}{^k} \, K \right)=\kappa \, S_{i}\,,
\end{equation}
\begin{equation} \label{YADMHamiltonianConstraint}
{^3}{R} + K^2 - K^{ij} \, K_{ij} = 2 \kappa \, \rho_m\,,
\end{equation}

\noindent where $\kappa=8\pi$ (setting $G=c=1$), $K_{ij}$ is the
extrinsic curvature, $\rho_m:=T_{tt}$ is the energy density defined with
respect to Gaussian normal observers, $S_{ij}=T_{ij}$ are the purely
spatial components of the energy-momentum tensor (with trace
$S:=\gamma^{ij} S_{ij}$), $S_i:=T_{ti}$ is the momentum density,
${^3}{R}_{ij}$ is the spatial Ricci tensor, and ${^3}{R}$ its trace.
Equations \eqref{YADMMomentumConstraint} and
\eqref{YADMHamiltonianConstraint} are constraints on the initial data,
and are referred to respectively as the momentum and Hamiltonian
constraints. The time evolution of the system is provided by
Eqs.~\eqref{YADMTimeEvolutionEquationsGNC1}
and~\eqref{YADMTimeEvolutionEquationsGNC2}.

First, we consider the constraints. The momentum constraint can in
principle be solved for the fluid velocity components $u^{\bar{\rho}}$
and $u^z$ but the constraint is quartic in $u^{\bar{\rho}}$ and $u^z$.
However, in the $\bar{\rho} \rightarrow 0$ limit, one component of the
momentum constraint equation reads:
\begin{equation} \label{MomentumConstraintUR}
\frac{\Delta z^2 \rho_u u^{\bar{\rho}} \chi^2
      \sqrt{\Delta z^2 (u^{\bar{\rho}})^2 \chi^2
            +\Delta z^2 \chi ^2 \left[(u^z)^2 \Psi +1\right]}
     }
     {
        \left(\Delta z^2 \chi ^2\right)^{3/2}
     } = 0,
\end{equation}

\noindent which implies $u^{\bar{\rho}}=0$ at $\bar{\rho}=0$. We
therefore require $u^{\bar{\rho}} \propto \bar{\rho}$. The Hamiltonian
constraint given in \eqref{YADMHamiltonianConstraint} can be solved for
the fluid density $\rho_u$ [see Eq. \eqref{RCM-AnisotropicFluidEMT}],
bearing in mind $\rho_m = T_{tt}$.

We now turn to the evolution equations. For the purposes of this
article, it suffices to compute the second time derivatives
$\ddot{\Psi}$, $\ddot{\chi}$, $\ddot{\lambda}$ at $t=0$ and
$\bar{\rho}=0$, given some specification for the initial data $\Psi$,
$\chi$, $\lambda$ and $\dot{\Psi}$, $\dot{\chi}$, $\dot{\lambda}$. At
the $t=0$ surface, the extrinsic curvature $K_{ij}$ may be computed by
taking the time derivative of $\gamma_{ij}$; its time derivative
$\partial_t K_{ij}$ may be computed similarly. One finds that each term
in Eq.~\eqref{YADMTimeEvolutionEquationsGNC2} has the same matrix form:
\begin{equation} \label{Matrixform}
M =
\left[
\begin{array}{ccc}
   a  &  0  &  b  \\
   0  &  0  &  0  \\
   b  &  0  &  c
\end{array}
\right],
\end{equation}

\noindent so that there are three independent equations for
$\ddot{\Psi}_0$, $\ddot{\chi}_0$, $\ddot{\lambda}_0$. In the $\bar{\rho}
\rightarrow 0$ limit, the equations yield the following
expressions (the details of the calculation are provided in the
associated \textit{Mathematica} file \cite{PortalMathematica2021}):
\begin{align}
  \ddot{\Psi}_0 &= \frac{\dot{\Psi}^2}{2 \Psi}, \nonumber\\
  \ddot{\chi}_0 &= \frac{\dot{\chi}}{2}
        \left[
              \frac{4 \dot{\chi}}{\chi} +
              \left(\frac{1}{\Psi }-\frac{4}{\Psi -1}\right) \dot{\Psi }
        \right]
        -\frac{{^3}R \Delta z \chi^2 \Psi u^z
        \partial_{\bar{\rho}} u^{\bar{\rho}}}
        {(\Psi-1)[1+(u^z)^2\Psi]}
        , \nonumber\\
  \ddot{\lambda}_0 &= \dot{\lambda }
        \left[
              \frac{4 \dot{\chi }}{\chi } -\frac{2 \dot{\Psi }}{\Psi -1}
              -\frac{(u^z)^2 \dot{\Psi }}{2 (u^z)^2 \Psi +2}
        \right] . \label{Solutionr}
\end{align}

These equations are subject to the condition:
\begin{equation} \label{Constraintl}
\dot{\lambda}_0 = \frac{\Delta z \chi \sqrt{-^{3}R \Psi} ~ u^z}
                  {\sqrt{2(\Psi -1) \left((u^z)^2 \Psi +1\right)}},
\end{equation}

\noindent which assumes $^{3}R<0$ (this is the case if $K$ and
$\rho_m<0$ dominate in the Hamiltonian constraint) and is needed to
ensure that $\ddot{\lambda}_0$ remains finite at $\bar{\rho}=0$. That an
additional constraint on $\dot{\lambda}_0$ is needed should not be
surprising, as we have already solved the Hamiltonian and momentum
constraints, and have fixed the gauge in choosing Gaussian normal
coordinates. It is well known that general relativity has two physical
degrees of freedom, and since there are three functions in the metric,
one might expect the equations to yield an additional constraint. We
note that if $\dot{\Psi}=\dot{\chi}=\dot{\lambda}=\Delta z=0$, Eq.
\eqref{Solutionr} implies $\ddot{\Psi}=\ddot{\chi}=\ddot{\lambda}=0$.
For the appropriate initial conditions at the core of the smoothed
string at $\bar{\rho}=0$, $\Delta z=0$, this ensures that no conical
singularity forms at the core of the string.

Now that we have an expression for $\ddot{\Psi}_0$, we can evaluate the
integrand for $A_0$. As it turns out, the solution
$\ddot{\Psi}_0=\frac{\dot{\Psi}_0^2}{2 \Psi_0}$ in Eq.~\eqref{Solutionr}
is precisely the condition for the vanishing of the integrand of $A_0$.
This leads us to the conclusion that for the class of portals we have
considered here, the portal mouths experience no acceleration toward
each other; the portal mouths neither attract nor repel, even when
brought close together.

%=======================================================================
%-----------------------------------------------------------------------
%
%		CURVATURE SINGULARITY FORMATION
%
%-----------------------------------------------------------------------
%=======================================================================
\subsection{Curvature singularity formation}

\begin{figure}
\includegraphics[width=0.95\columnwidth]{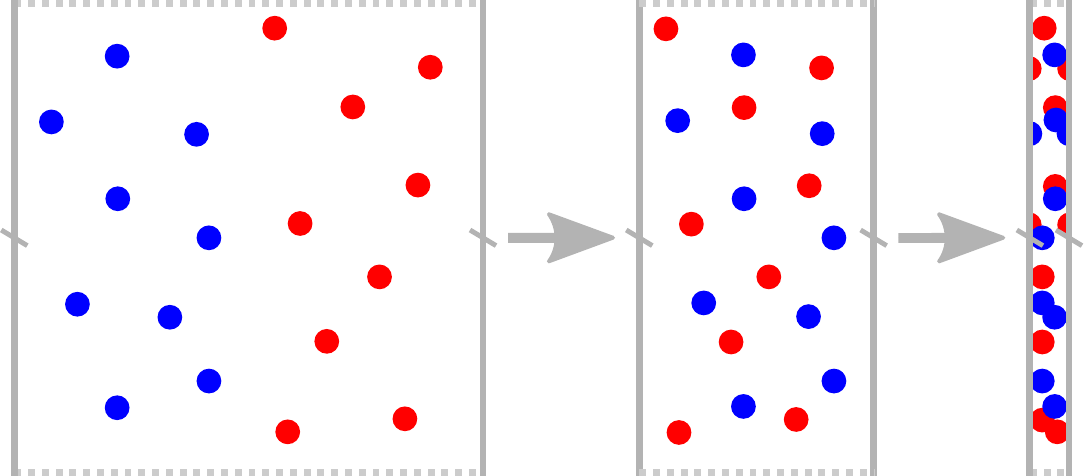}
\caption{An illustration describing the behavior of a rigid arrangement
of cosmic strings between the portal mouths as the mouths approach each
other (the leftmost diagram is earliest in time). Here, the $z$
direction is horizontal, the radial direction is vertical, and the
strings (indicated by the red and blue dots) are perpendicular to the
page and parallel to the $\phi$ direction. The portal mouths correspond
to the identified vertical lines. As the distance between the portal
mouths decreases, the density of cosmic strings increases.}
\label{fig:Density}
\end{figure}

Since the absence of an effective force between the portal mouths
indicates that classical general relativity presents no obstruction to
their collision, one might conceivably imagine that the collision
process results in topology change. A topology changing process will
likely require the tearing of the manifold, and one might expect the
formation of a curvature singularity as the portal mouths are brought
together.

To see that a curvature singularity does indeed form, we analyze what
happens to the Ricci scalar as the portal mouths are brought together.
Consider the case where the portal mouths are approaching each other
with some initial velocity given by some specification of initial data
for $\dot{\Psi}$. One might imagine that since
Eq.~\eqref{RCM-LineElemCylCoordsSmoothedTimeDep} has a rather general
form, the metric at a later time may be recast into the same form with a
smaller value of $z_0$ and $\chi \sim 1$ at $\bar{\rho}=0$. If in this
case $\Psi$ is of order unity, a significantly smaller value for $z_0$
will correspond to a decreased proper distance $L$ [as defined in
Eq.~\eqref{ProperDistance}] between the portal mouths. At
$\bar{\rho}=0$, the Ricci scalar takes the form:
\begin{equation} \label{RCM-RicciLine2}
  \begin{aligned}
    R |_{{\bar{\rho}}=0} &= \frac{2 \ddot{\Psi} \Psi - \dot{\Psi}^2
    - 2 \Psi \partial_{\bar{\rho}} \partial_{\bar{\rho}} \Psi }
    {2 \Psi ^2} \\
    &
    +
    \frac{ (\Psi +1) \chi \partial_z \Psi -
    2 (\Psi -1) \Psi \partial_z \chi }{\Delta z \chi ^2 \Psi ^2}\\
    &
    -\frac{2 (\Psi -1) \left(\partial_z \lambda^2+\chi -1 -
    \Psi \dot{\lambda}^2 \right)}{\Delta z^2 \chi ^2 \Psi }.
  \end{aligned}
\end{equation}

\noindent For points $|\Delta z|<z_0$, the last two terms in the Ricci
scalar become large as $z_0 \rightarrow 0$. It is straightforward to
show that $\partial_z \Psi \sim \frac{k_z}{z_0}$, with $k_z$ of order
unity. At the core of the string $\bar{\rho} =0$, $\Delta z =0$, one
must have $\Psi=\alpha^2=4$ to ensure that the spatial geometry there
remains locally flat. Though the boundary conditions for satisfying the
junction conditions at $z=0$ for the line element
\eqref{RCM-LineElemCylCoordsSmoothedTimeDep} are more complicated, one
can still establish that the junction conditions require $\Psi=1$ at at
$\bar{\rho}=0$. Since $\Psi=4$ at $z=z_0$ and $\Psi=1$ at $z=0$ along
$\bar{\rho}=0$, one concludes that for some value of $|\Delta z|<z_0$,
$\partial_z \Psi \sim \frac{k_z}{z_0}$, with $k_z$ of order unity.

Now at $z=0$, one might also expect $z$ derivatives to vanish since the
geometry is symmetric about $z=0$. Since $\Psi=1$ at $\bar{\rho}=0$, and
assuming that the dynamics preserve the symmetries
($\ddot{\Psi}=\dot{\Psi}=0$), the Ricci scalar simplifies to:
\begin{equation} \label{RCM-RicciLine2sym}
R |_{\bar{\rho}=z=0} = - \partial_{\bar{\rho}} \partial_{\bar{\rho}}
                         \Psi.
\end{equation}

\noindent In principle, the Ricci scalar at $\bar{\rho}=0, z=0$ may
diverge in the $z_0 \rightarrow 0$ limit if $\partial_{\bar{\rho}}
\partial_{\bar{\rho}} \Psi$ diverges.

Intuitively, the formation of a singularity in the Ricci curvature
indicates that the density of anisotropic fluid becomes large as the
portal mouths approach each other. To understand this, consider what
happens to the matter between the portal mouths in the region
$\bar{\rho}<a$, $|\Delta z|<z_0$ for finite $a$. In between the portal
mouths, curves parallel to the $z$-axis (by which we mean curves of
constant $t, \bar{\rho}, \phi$) have finite length on the order of the
separation distance $z_0 \sim L$ between the portal mouths. As the
portal mouths approach each other, the separation distance decreases,
and it follows that the volume of the region between the portal mouths
decreases. For smoothed cosmic strings, there will always be a finite
amount of matter in the region $\bar{\rho}<a$, $|\Delta z|<z_0$, and it
follows that the density of the matter must become large as the portal
mouths approach each other. This mechanism is illustrated in Fig.
\ref{fig:Density}, which depicts the anisotropic matter distribution
between the portal mouths as a bundle of (low) negative mass cosmic
strings.

%=======================================================================

%-----------------------------------------------------------------------
%-----------------------------------
%-----------------
%--------
%---
%-
%
%
%-
%---
%--------
%-----------------
%-----------------------------------
%-----------------------------------------------------------------------
%=======================================================================
%-----------------------------------------------------------------------
%
%		CONCLUSION
%
%-----------------------------------------------------------------------
%=======================================================================
\section{Conclusion}
\label{conc}

We have described in detail the construction and topology of portals
supported by smoothed matter distributions, and have obtained line
elements, given in Eqs.~\eqref{RCM-LineElemCylCoordsSmoothed} and
\eqref{RCM-LineElemCylCoordsSmoothedTimeDep} for a class of (large
radius) axisymmetric portals supported by an anisotropic fluid. We have
shown that the portal mouths experience no acceleration toward each
other, which in turn suggests that there is no effective force present
between the portal mouths in the direction parallel to the axis of
symmetry. We have also shown that the Ricci scalar diverges as the
portal mouths collide, as one might expect. These results indicate that
for the class of portal-type geometries considered, gravity alone does
not prevent a collision between mouths of smoothed portals, and that,
since the Ricci scalar diverges, a complete description of such a
collision and the final state will likely require a theory of quantum
gravity. Regardless, one might be able to make some progress toward
understanding the aftermath of portal self-collisions by relaxing
symmetry assumptions and considering the interactions of unsmoothed
cosmic strings. The interaction of unsmoothed cosmic strings as
described in \cite{Hellaby1991,Ellis1992} suggests that in general, the
self-collision of asymmetric portal geometries in such a case will
likely result in highly nontrivial spatial topologies and cosmic string
configurations.

Our results are based on a simple classical matter model, that of an
anisotropic fluid, in which the stresses for the smoothed cosmic string
are simplified in that they are directed only along the length of the
string, at least in the large $a$ limit we have considered. Since there
are no stresses in the directions perpendicular to the strings, the
fluid does not contribute to the effective force between portal mouths.
The lack of acceleration between the portal mouths arising from the
solution to the Einstein equations indicates that gravity does not
generate an effective force between the portal mouths. One might expect
this to be the case if one imagines the anisotropic fluid to consist of
a bundle of low (negative) mass/tension cosmic strings; since the
geometry in the immediate region around a single unsmoothed cosmic
string is locally flat, a collection of strings will not gravitate.

One might expect our results to change when quantum effects are
included. In the region between the portal mouths $\rho<a$, $|\Delta
z|<z_0$, curves parallel to the $z$-axis have topology $S^1$, which
suggests that quantum fields within this region must satisfy periodic
boundary conditions. One may then expect the portal mouths to experience
an attractive (topological) Casimir force when the portal mouths are
separated by small distances (see 12.3.3 of \cite{Visser1995} and also
\cite{Fulling:1989nb} for a discussion of the Casimir effect with
periodic boundary conditions).

A more detailed matter model for the smoothed cosmic string could also
produce an effective force between the portal mouths, changing our
result. One might, for instance, consider an electrically charged cosmic
string, which would produce a repulsion between the portal mouths if the
overall charge is nonzero, or a neutral current-carrying string, which
would attract, as the symmetry about $z=0$ implies that the currents are
parallel. It is also worth investigating whether strings with negative
mass can be constructed in null energy condition-violating field
theories, such as the Einstein-Dirac-Maxwell theory. A simple example
involves the construction of negative mass strings from phantom fields,
in which the action for the matter sector for the fields has the
opposite sign relative to the gravitational action. In the case of local
gauge strings \cite{Garfinkle1985,Laguna-Castillo1987}, one might
imagine that in the long-distance limit (in which gravity is weak),
their phantom counterparts interact similarly, as the actions differ
only by an overall sign. From numerical studies of interacting local
gauge strings \cite{Matzner1988}, parallel oriented strings weakly
repel, so one might expect a weak repulsion for phantom gauge strings.

These considerations indicate that the nature and strength of the
interaction between portal mouths are not determined primarily by the
gravitational interaction (as described by classical general
relativity). Rather, the behavior of closely separated portal mouths
depends critically on the behavior of matter---both the matter
supporting the portals and the behavior of quantum fields around portals
can change the direction and magnitude of the effective force between
the portal mouths. An interesting question, motivated by the singularity
which forms as the portal mouths are brought together, is whether
low-energy quantum gravitational effects, manifesting as
higher-curvature terms in the effective action, introduce nontrivial
gravitational interactions between the portal mouths.

%=======================================================================

%-----------------------------------------------------------------------
%-----------------------------------
%-----------------
%--------
%---
%-
%
%
%-
%---
%--------
%-----------------
%-----------------------------------
%-----------------------------------------------------------------------

%=======================================================================
%		ACKNOWLEDGMENTS
%=======================================================================

\begin{acknowledgments}
We thank Mark Baumann for helpful discussions. J.C.F. acknowledges
financial support from FCT - Funda\c c\~ao para a Ci\^encia e Tecnologia
of Portugal Grant No.~PTDC/MAT-APL/30043/2017 and Project
No.~UIDB/00099/2020.
\end{acknowledgments}

%=======================================================================
%		BIBLIOGRAPHY
%=======================================================================

%

%=======================================================================

\end{document}